# Migration of D-type asteroids from the outer Solar System inferred from carbonate in meteorites

**Authors:** W. Fujiya[1]*, P. Hoppe[2], T. Ushikubo[3], K. Fukuda[4]†, P. Lindgren[5], M. R. Lee[6], M. Koike[7]‡, K. Shirai[8]§, and Y. Sano[7]

**Affiliations:**

[1]Faculty of Science, Ibaraki University, 2-1-1 Bunkyo, Mito, Ibaraki 310-8512, Japan.

[2]Max Planck Institute for Chemistry, Hahn-Meitner-Weg 1, 55128 Mainz, Germany.

[3]Kochi Institute for Core Sample Research, Japan Agency for Marine-Earth Science and Technology (JAMSTEC), 200 Monobe-otsu, Nankoku, Kochi 783-8502, Japan.

[4]Department of Earth and Planetary Science, The University of Tokyo, 7-3-1 Hongo, Bunkyo, 113-0033 Tokyo, Japan.

[5]Department of Geology, Lund University, Sölvegatan 12, 223 62 Lund, Sweden.

[6]School of Geographical & Earth Sciences, University of Glasgow, Gregory Building, Lilybank Gardens, Glasgow G12 8QQ, UK.

[7]Atmosphere and Ocean Research Institute, The University of Tokyo, 5-1-5 Kashiwanoha, Kashiwa, Chiba 277-8564, Japan.

[8]International Coastal Research Center, Atmosphere and Ocean Research Institute, The University of Tokyo, The University of Tokyo, 2-106-1 Akahama, Otsuchi, Iwate 028-1102, Japan.

*Correspondence to: wataru.fujiya.sci@vc.ibaraki.ac.jp

†Current affiliation: Department of Geoscience, University of Wisconsin-Madison, 1215 W. Dayton St., Madison WI 53706, USA.

‡Current affiliation: Department of Solar System Science, Institute of Space and Astronautical Science (ISAS), Japan Aerospace Exploration Agency (JAXA), 3-1-1 Yoshinodai, Chuo-ku, Sagamihara, Kanagawa 252-5210, Japan.

§Current affiliation: Atmosphere and Ocean Research Institute, The University of Tokyo, 5-1-5 Kashiwanoha, Kashiwa, Chiba 277-8564, Japan.

**Recent dynamical models of Solar System evolution and isotope studies of rock-forming elements in meteorites have suggested that volatile-rich asteroids formed in the outer Solar System beyond Jupiter's orbit, despite being currently located in the main asteroid belt[1-4]. The ambient temperature under which asteroids formed is a crucial diagnostic to pinpoint the original location of asteroids and is potentially determined by the abundance of volatiles they contain. In particular, abundances and $^{13}C/^{12}C$ ratios of carbonates in meteorites record the abundances of carbon-bearing volatile species in their parent asteroids. However, the sources of carbon for these carbonates remain poorly understood[5-8]. Here we show that the Tagish Lake meteorite contains abundant carbonates with consistently high $^{13}C/^{12}C$ ratios. The high abundance of $^{13}C$-rich carbonates in Tagish Lake excludes organic matter as their main carbon source[5,9]. Therefore, the Tagish Lake parent body, presumably a D-type asteroid[10], must have accreted a large amount of $^{13}C$-rich $CO_2$ ice. The estimated $^{13}C/^{12}C$ and $CO_2/H_2O$ ratios of ice in Tagish Lake are similar to those of cometary ice[11,12]. Thus, we infer that at least some D-type asteroids formed in the cold outer Solar System and were subsequently transported into the inner Solar System owing to an orbital instability of the giant planets[1,3].**

We performed in situ C-isotope measurements on individual grains of carbonate minerals, calcite ($CaCO_3$) and dolomite ($CaMg(CO_3)_2$), in Tagish Lake, an ungrouped carbonaceous chondrite (CC) with a petrologic type of 2, indicating that it underwent aqueous alteration[13,14] (Methods). For comparison, we also conducted C- and O-isotope measurements on calcite grains in two Mighei-type carbonaceous chondrites (CM chondrites) with a petrologic type of 2, Nogoya and LaPaz Icefield (LAP) 031166. Due to the small grain size of Tagish Lake carbonates, we were not able to measure O isotopic ratios (Supplementary Fig. 1).

The $\delta^{13}C$ and $\delta^{18}O$ values ($^{13}C/^{12}C$ and $^{17,18}O/^{16}O$ ratios are also expressed as $\delta^{13}C_{VPDB}$ and $\delta^{17,18}O_{VSMOW}$, respectively, which represent permil, $10^{-3}$ expressed as ‰, deviations from the isotopic ratios of standard materials: VPDB, Vienna Pee Dee Belemnite; VSMOW, Vienna Standard Mean Ocean Water) of CM carbonates are highly variable, ranging from approximately 20 to 80‰ and from approximately 15 to 40‰, respectively, and do not correlate with each other (Fig. 1a and Supplementary Table 1). The O-isotope data of CM carbonates plot on a single trend line (Fig. 1b), reflecting a change in formation temperatures and/or in the O isotopic ratios of fluids from which they formed. Therefore, the lack of correlation between $\delta^{13}C$ and $\delta^{18}O$ indicates that the variable $\delta^{13}C$ values of CM carbonates must reflect isotopic heterogeneity of carbon sources (Methods). In contrast to CM carbonates, Tagish Lake carbonates have consistently high $\delta^{13}C$ values of approximately 70‰ (Fig. 2 and Supplementary Table 2), which indicates the presence of a dominant carbon reservoir with a high $\delta^{13}C$ value in the fluid. Thus, the distinct $\delta^{13}C$ characteristics of CM and Tagish Lake carbonates suggest that at least two carbon reservoirs, one with high $\delta^{13}C$ (>80‰) and another with low $\delta^{13}C$ (<20‰) values, were accreted to the parent bodies in different mixing ratios (see Supplementary Information for arguments against other proposed scenarios).

Possible candidates of such carbon reservoirs would be $CO_2$, CO and $CH_4$ ices and organic matter. $CO_2$ is the major C-bearing molecule in comets[15]. If organic matter contributed carbon to the carbonates, then it must have been first oxidized to produce $CO_2$. Oxidation of organic matter requires strong oxidants such as peroxide, possibly contained in ice[9]. The maximum amount of $CO_2$ that could have been produced from organic matter in CCs is approximately 0.1 wt% (ref. [5]), which is far lower than the amount of carbon in Tagish Lake carbonates (approximately 1.3 wt%; represented by weight percent of the carbonate carbon)[16]. Furthermore, the highest $\delta^{13}C$ value observed for soluble organic compounds is approximately 60‰ (ref. [17]), which is lower than the highest $\delta^{13}C$ value of approximately 80‰ observed for CM carbonates. Therefore, given the consistently high $\delta^{13}C$ values of the Tagish Lake carbonates, the source of high $\delta^{13}C$ carbon was almost certainly $^{13}C$-rich $CO_2$ ice, which, after sublimation, dissolved in water to form the carbonates during parent-body aqueous alteration. This conclusion is consistent with the $\delta^{13}C$ value of 65 ± 51‰ for $CO_2$ in the coma of comet 67P/Churyumov-Gerasimenko recently measured by the Rosetta spacecraft[12]. In contrast, the carbon reservoir(s) with a low $\delta^{13}C$ value are poorly constrained, although organic matter seems most likely (see Supplementary Information).

For the calculations below, we assume that the $\delta^{13}C$ values of $CO_2$ ice and the other as yet unidentified C reservoir are 80‰ and 20‰, respectively. We note that the uncertainty on these endmember compositions does not affect our conclusions (Methods). The average $\delta^{13}C$ value of CM carbonates, as obtained from whole-rock samples, is approximately 45‰ (ref. [5]), which indicates that approximately 42% of the carbonate carbon would be derived from $CO_2$ ice based on a mass-balance calculation (Supplementary Table 3). Thus, the variable $\delta^{13}C$ values of CM carbonates likely reflect an incomplete mixing between carbon from $CO_2$ ice and another $^{13}C$-poor reservoir, which were accreted to the CM parent body in roughly similar proportions. In contrast, the high abundance and consistently high $\delta^{13}C$ values of Tagish Lake carbonates suggest that its parent body contained $CO_2$ ice that was sufficiently abundant to swamp carbon from other reservoirs. Given the whole-rock $\delta^{13}C$ value (approximately 68‰) of the Tagish Lake carbonate[16], as much as approximately 80% of the carbonate carbon would have been derived from $CO_2$ ice (Supplementary Table 3).

Using the values above and the $H_2O$ abundances of CM chondrites and Tagish Lake, one can calculate the $CO_2/H_2O$ mole ratios of ice accreted to the CM and Tagish Lake parent bodies. The $H_2O$ abundance of water-rich CCs can be estimated from the hydrogen contents of hydrous minerals[18,19] (also see Methods). The inferred $CO_2/H_2O$ ratios of CM ice range from approximately 0.001 to 0.060 (0.012 on average). In contrast, the inferred $CO_2/H_2O$ ratio of the Tagish Lake ice (approximately 0.24) is much higher (Supplementary Table 3), and lies within the cometary range (Fig. 3) (0.04 to 0.30, average 0.16) (ref. [11]). These findings clearly demonstrate that the composition of ice accreted to the Tagish Lake parent body is distinct from the CM parent body, and is similar to cometary ice.

Although no in situ C-isotope data of individual carbonate grains in Ivuna-type (CI) and Renazzo-type (CR) chondrites are available, the whole-rock isotopic ratio of carbonate carbon in CRs has a wide range[5], in common with CM carbonates. For CI carbonates, the variability of whole-rock $\delta^{13}C$ values is poorly constrained, because of limited statistics[5]. Furthermore, the carbonate carbon abundances in CI and CR chondrites are much lower than in Tagish Lake[5]. These observations may suggest that the CI and CR parent bodies also accreted smaller amounts of $CO_2$ ice than the Tagish Lake parent body, and they are genetically closer connected to the CM parent body.

A dichotomy in the ratios of non-traditional stable isotopes of, for example, chromium, titanium, molybdenum and tungsten, between CCs and non-carbonaceous meteorites suggests that the CC and non-carbonaceous meteorite reservoirs were spatially separated by Jupiter and that the CC parent bodies formed beyond the orbit of Jupiter, that is, at >5 au (ref. [4]). Most CCs, including CMs, have been linked to C-type asteroids from their reflectance spectra[20]. The condensation temperature of $CO_2$ in the solar nebula is approximately 80 K (ref. [21]), and the CM parent body (C-type asteroid) likely formed sunward of the $CO_2$ condensation front. The precise temperature and distance from the Sun at which C-type asteroids formed are, however, difficult to determine because some $CO_2$ may have been trapped in $H_2O$ ice even at temperatures higher than 80 K (ref. [21]). Other complications may be local temperature fields that were likely present in the solar nebula, especially around infant Jupiter[22].

In contrast to the CM parent body, the Tagish Lake parent body formed where substantial amounts of $CO_2$ could have condensed to ice, in common with comets, and mostly likely beyond the $CO_2$ condensation front. The reflectance spectrum of Tagish Lake suggests that it is derived from a D-type asteroid[10]. Assuming that the $CO_2$ condensation front in the solar nebula was at a distance of >10 au from the Sun[23], at least some D-type asteroids accreted in the cold Outer Solar system where the ice giant planets formed, or even farther out, for example, in the trans-Neptunian regions. D-type asteroids occur mainly in the outer edge of the main asteroid belt (2.1-3.3 au) and in the Jupiter Trojan regions[24]. The D-type asteroids that today occur in the main belt and the Trojan regions must have subsequently migrated inwards owing to an orbital instability of giant planets possibly as late as about 4.1 billion years ago[1,3]. Tagish Lake contains organic globules, which resemble cometary carbon, hydrogen, oxygen and nitrogen (CHON) particles, with isotopic compositions being inherited from cold molecular clouds[25]. The low density and high carbon content of Tagish Lake also testify its nature as a primitive meteorite that formed in a cold environment[13,16].

Our conclusion might be inconsistent with the H isotopic ratio (D/H ratio) of Tagish Lake water (and CC water in general), which is much lower than that of cometary ice[18]. However, we note that the D/H ratio of the Tagish Lake water is poorly constrained compared with water in CM and CR chondrites owing to the large uncertainty of the mass-balance calculation attributed to the limited number of measurements[18,26]. Furthermore, the Jupiter-family comet 103P/Hartley 2 has a low D/H ratio in common with CC water[27], demonstrating considerable variation in the D/H ratios of cometary ices within which the Tagish Lake water may lie.

Debates around where asteroids formed are central to longstanding arguments concerning the sources of volatiles for the inner planets. Our results suggest that materials from the cold outer Solar System were supplied to the asteroid belt, that is, the inner Solar System, although the total mass of D-type asteroids is relatively small[24]. CC-like material was an important constituent of planets in the inner Solar System, including Earth, with contributions of a few percent to the total mass[28], especially for volatiles. Thus, a heritage from the outer Solar System may be present on the Earth, albeit in limited quantities. The concentrations and isotopic ratios of volatiles in comet 67P/Churyumov-Gerasimenko suggest that the present-day Earth's atmosphere contains cometary noble gas (for example, ~20% of the total atmospheric xenon), while the cometary contribution to terrestrial carbon appears lower than 1% (refs. [29,30]). Future exploration of D-type asteroids as well as further analysis of Tagish Lake will unravel the evolution of small icy bodies and reveal the extent to which these bodies contributed to formation of the inner planets.

**Correspondence and requests for materials** should be addressed to W. F.

**Acknowledgments**

We thank Naoto Takahata, Takanori Kagoshima, Akizumi Ishida, and Elmar Gröner for assistance of the ion probe analyses. This work was supported by JSPS KAKENHI Grant Numbers 16K17838, 17K18814 and 18H04454, and UK Science and Technology Facilities Council (STFC) grant ST/N000846/1.

**Author contributions**

W. F. designed this work, W. F., P. H., U. T., K. F., M. K., and Y. S. performed the ion probe analyses, P. L. and M. R. L. carried out the petrologic and mineralogical observations of the samples, K. S. prepared the standard materials for the ion probe analyses, and all authors participated in discussion and preparation of the manuscript.

**Competing interests**

Authors declare no competing interests.

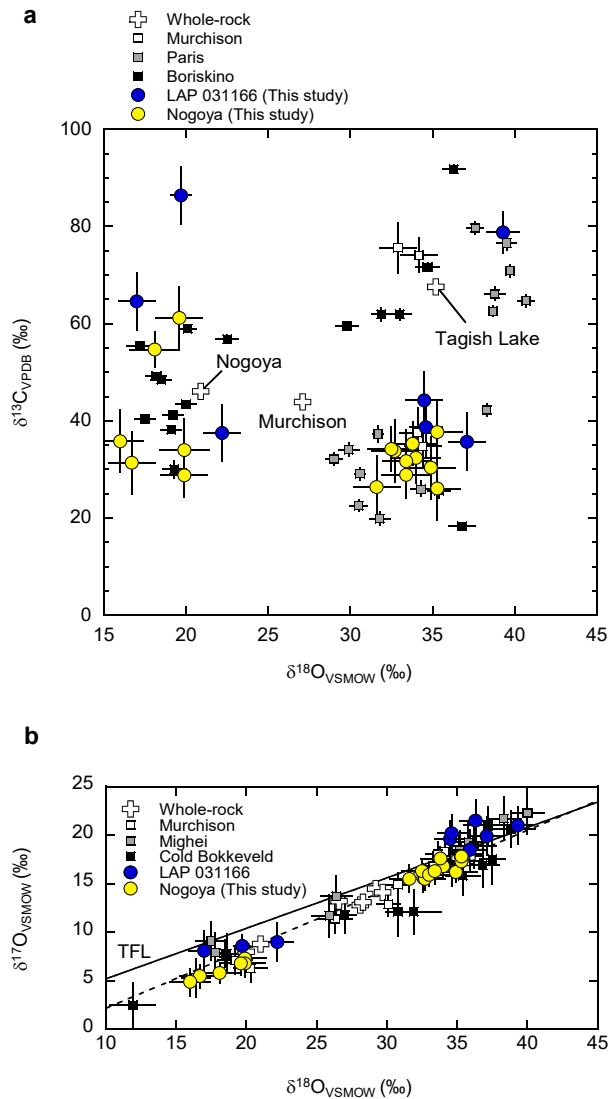

**Fig. 1 | C and O isotopic ratios of Ca-carbonate grains in the CM chondrites Nogoya and LAP 031166. a**, $\delta^{13}C_{VPDB}$ versus $\delta^{18}O_{VSMOW}$ values. The delta notation represents a permil deviation from a standard material: VPDB, Vienna Pee Dee Belemnite; VSMOW, Vienna Standard Mean Ocean Water. Also shown are the data of individual Ca-carbonate grains in the CM chondrites Murchison, Paris and Boriskino, and the whole-rock data of Murchison, Nogoya and Tagish Lake from the literature[5-8]. The CM carbonates show a large variation in $\delta^{13}C$ ranging from approximately 20 to 80‰, not correlated with $\delta^{18}O$. **b**, $\delta^{17}O_{VSMOW}$ versus $\delta^{18}O_{VSMOW}$ values. Also shown are the data of individual Ca-carbonate grains in the CM chondrites Murchison, Mighei and Cold Bokkeveld, and those obtained for whole-rock CM chondrites and their regression line (the dashed line) from the literature[38,48]. TFL, terrestrial fractionation line. Errors are the external reproducibility (2 standard deviation) of repeated measurements on the standard materials.

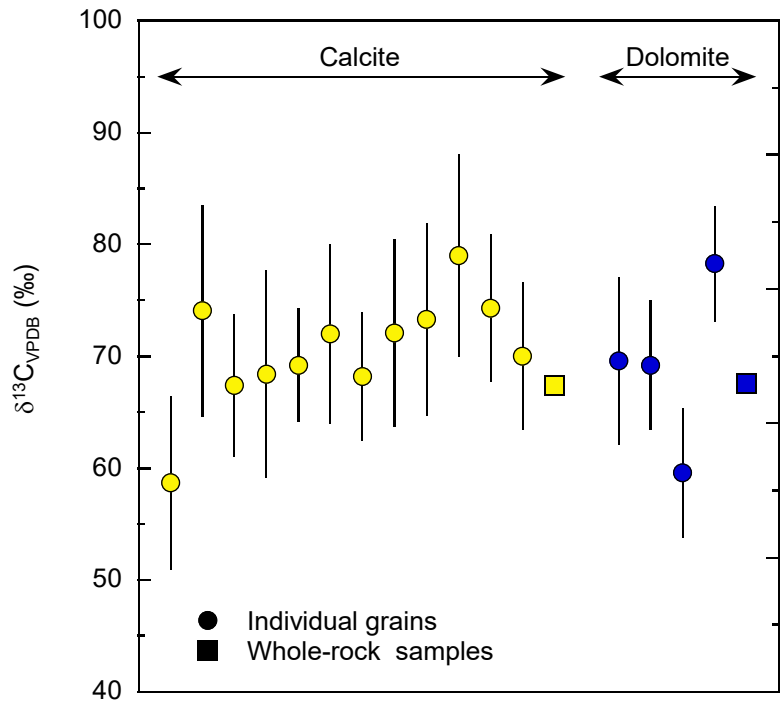

**Fig. 2 | $\delta^{13}C_{VPDB}$ values of calcite and dolomite grains in Tagish Lake.** Also shown are those measured for whole-rock samples[16]. The Tagish Lake carbonates have consistently high $\delta^{13}C$ values of approximately 70‰. The average $\delta^{13}C$ values of the individual grains are indistinguishable from those obtained by analysis of whole-rock samples. Errors (2 sigma) are the external reproducibility of repeated measurements on the standard materials or the internal precision from counting statistics, whichever is larger.

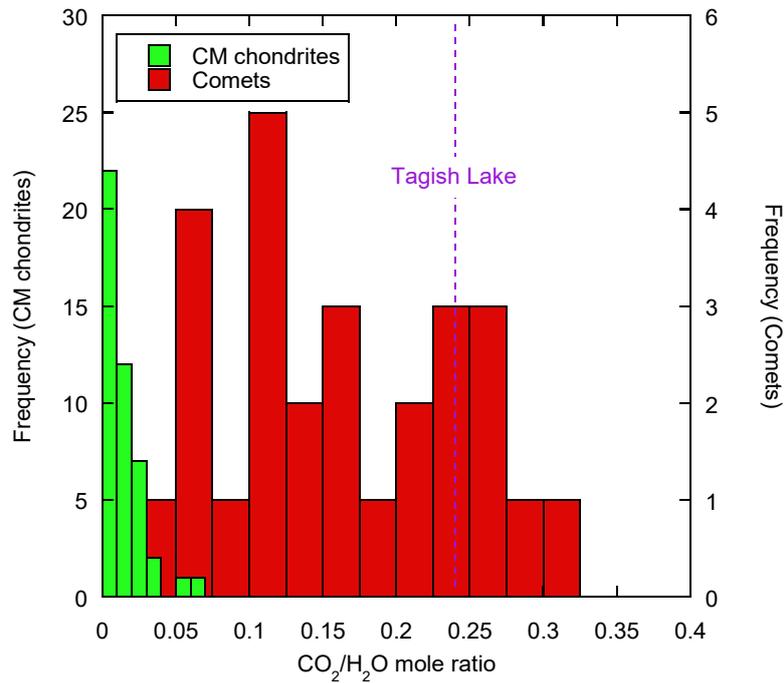

**Fig. 3 | Histogram of the $CO_2/H_2O$ mole ratios of ice in CM chondrites and comets.** The $CO_2/H_2O$ ratios of ice in individual CM chondrites are calculated from carbonate abundances and hydrogen contents[5,19] (see Methods). Data for comets are from the literature[11]. The dashed line represents the $CO_2/H_2O$ mole ratio of ice in Tagish Lake (~0.24). $CO_2$ ice abundances of comets are higher than those of CMs in general, whereas that of Tagish Lake is within the cometary range.

**Methods**

**Samples.** Mineralogical and petrological observations were performed on polished thin sections of Nogoya, LAP 031166 and Tagish Lake with scanning electron microscopes at Ibaraki University (JEOL JSM-5600LV) and at University of Glasgow (FEI Quanta 200F and Zeiss Sigma). Carbonate grains were identified using energy dispersive X-ray spectroscopy equipped with the scanning electron microscope.

Nogoya is a moderately altered CM chondrite with a petrographic type of 2.3 (ref. [31]). Significant amounts, but not all, of anhydrous silicate are aqueously altered to phyllosilicate. Carbonate minerals in Nogoya are predominantly calcite ($CaCO_3$) and are dispersed throughout the matrix (Supplementary Fig. 1). Rare occurrence of dolomite ($CaMg(CO_3)_2$) has been reported previously[32-35], but it was not found in this study. The carbonate abundance of Nogoya, represented by weight percent of the carbonate carbon, is approximately 0.10 wt% (ref. [5]). Notably, we identified two types of calcite, namely, type 1 and 2, in terms of their morphology[35-37]. Type 1 calcite occurs as small equant grains and precipitated in pore water. Most type 1 grains are rimmed by tochilinite and serpentine. Type 2 calcite is commonly porous with tiny inclusions and replaced other minerals such as silicate. In Nogoya, type 1 calcite is more common than type 2. O and C isotopic ratios of twelve type 1 grains and six type 2 grains were analysed with secondary ion mass spectrometry (SIMS).

LAP 031166 is a more aqueously altered CM chondrite than Nogoya. Its petrological type is 2.1 (ref. [35]) and most anhydrous silicate is replaced by serpentine. Carbonate can be found throughout the matrix. The only carbonate mineral found in LAP 031166 is calcite (Supplementary Fig. 1). This carbonate occurrence is contrary to the observation of other CM 2.1 chondrites, for example, Allan Hills 83100 and Queen Alexandra Range 93005, where calcite as well as dolomite are common[35]. Contrary to Nogoya, LAP 031166 contains comparable amounts of type 1 and 2 calcites. The O isotopic ratios of calcite grains were reported previously[38]. Here, we measured C isotopic ratios of the four type 1 and three type 2 calcite grains, whose O isotopic ratios were already known.

Tagish Lake is an ungrouped, heavily altered carbonaceous chondrite[13,14]. Its reflectance spectrum resembles those of D-type asteroids and does not match to any other chondrites[10,39]. There are sparse chondrules where most anhydrous silicate is replaced by phyllosilicate. Carbonate minerals prevail in this meteorite. It should be noted that Tagish Lake has two lithologies, carbonate-poor and carbonate-rich, where the dominant carbonate species are different[14]. In the carbonate-poor lithology, calcite is more common than other carbonate minerals such as dolomite (Supplementary Fig. 1). The calcite in the carbonate-poor lithology is typically smaller than approximately 10 µm, which makes ion probe analyses difficult. In the carbonate-rich lithology, Fe-Mg-Ca-Mn carbonate is the dominant carbonate mineral and is very abundant. The Fe-Mg-Ca-Mn carbonate is typically less than a few micrometres, even smaller than the calcite in the carbonate-poor lithology. In this study, we measured C isotopic ratios of twelve calcite grains as well as four dolomite grains in the carbonate-poor lithology.

In a previous study, the abundance and C isotopic ratio of carbonate in Tagish Lake were measured for two whole-rock samples[16]. The two results are very similar and suggest a high carbonate abundance of approximately 1.3 wt% represented by weight percent of the carbonate carbon. Assuming that calcium in Tagish Lake with an amount of 0.99 wt% (ref. [13]) was entirely consumed to produce calcite, one can obtain a calcite abundance of approximately 0.30 (=0.99×12/40) wt%. This abundance sets the upper limit on the carbonate abundance in the carbonate-poor lithology, because not all calcium will be transformed to calcite, which is by far the dominant carbonate mineral in this lithology. This upper limit is much smaller than the measured overall carbonate abundance of approximately 1.3 wt%. In the carbonate-rich lithology, the matrix accounts for 83 vol% and relative abundances of phyllosilicate and carbonate in the matrix is roughly 3:2 (ref. [40]). Thus, given the density of Tagish Lake (~1.7 g cm$^{-3}$) and siderite ($FeCO_3$: ~4.0 g cm$^{-3}$), and the porosity of Tagish Lake (37 %) (ref. [41]), the carbonate abundance in the carbonate-rich lithology would be roughly 5.1 (=0.83×0.63×2/5×12/116×4.0/1.7×100) wt%, larger than the measured carbonate abundance (here we assume that siderite is the only carbonate mineral in the carbonate-rich lithology). These observations indicate that the whole-rock samples analysed previously were composed of both carbonate-poor and -rich lithologies. The above conclusion is supported by the fact that the whole-rock samples contain both Ca- and Fe,Mg-carbonates[16]. It should be stressed that the $\delta^{13}C$ values of the calcite and dolomite measured for the two whole-rock samples (67.4 ‰ and 67.6 ‰, respectively[16]) are essentially the same as the average $\delta^{13}C$ values of the individual calcite and dolomite grains (70.6±2.9‰ and 69.2±7.6‰, respectively) obtained in this study, which indicates that the consistently high $\delta^{13}C$ values of individual grains cannot be explained by a sampling bias. Although we were not able to analyse Fe-Mg-Ca-Mn carbonate grains in the carbonate-rich lithology due to the small grain size, it is likely that they also have homogeneous $\delta^{13}C$ values consistent with the carbonate grains in the carbonate-poor lithology measured in this study.

**Isotope measurements.** O isotopic ratios of calcite grains in Nogoya were measured with the CAMECA IMS 1280-HR at Kochi Institute for Core Sample Research, Japan Agency for Marine-Earth Science and Technology. We used a terrestrial calcite standard, UWC-3, from the Adirondack Mountains of New York with a $\delta^{18}O_{VSMOW}$ value of 12.49‰ (ref. [42]). Negative secondary ions of $^{16}O^-$, $^{17}O^-$ and $^{18}O^-$ were produced with a focused $Cs^+$ primary ion beam (~30 pA, 3-4 µm in diameter). A Faraday cup and two electron multipliers were used to detect $^{16}O^-$ and $^{17,18}O^-$ ions, respectively. An electron gun was utilized for charge compensation. The typical count rate of $^{16}O^-$ was $2\times10^7$-$4\times10^7$ cps and the dead time correction of the electron multipliers was applied. The measurement time was approximately 10 min including presputtering. The contribution from the tail of $^{16}OH^-$ to $^{17}O^-$ mass spectra was estimated to be <0.07‰ (for the standard calcite) and 0.02-0.7‰ (for the meteoritic calcite, depending on the purity of the samples). The external reproducibility (2σ) of the analyses was typically ~1.4, ~1.2 and ~0.9‰ for $\delta^{18}O$, $\delta^{17}O$ and $\Delta^{17}O$ values, respectively ($\Delta^{17}O$ represents deviation from the terrestrial fractionation line: defined by $\Delta^{17}O=\delta^{17}O-0.52\times\delta^{18}O$). Details of the data reduction are described in the literature[43,44]. Note that we previously measured the O isotopic ratios of the same calcite grains with a NanoSIMS 50 (ref. [45]). The data obtained with the NanoSIMS 50 and IMS 1280-HR are generally in good agreement, but the analytical precision of the IMS 1280-HR analyses was much better than that of the NanoSIMS 50 analyses. Therefore, we adopted the data obtained with the IMS 1280-HR for discussion.

C-isotope analyses on the calcite grains in Nogoya and LAP 031166 were performed with the CAMECA NanoSIMS 50 at Atmosphere and Ocean Research Institute, The University of Tokyo. Negative secondary ions of $^{12}C^-$, $^{13}C^-$, $^{18}O^-$, $^{12}C^{14}N^-$ and $^{28}Si^-$, produced by rastering a $Cs^+$ primary ion beam (20-30 pA, ~1 µm in diameter) over $6\times6$ µm$^2$-sized areas, were detected simultaneously with five electron multipliers. The measurement time was about 12 min including presputtering. A terrestrial calcite rock from Mexico with a $\delta^{13}C_{VPDB}$ value of 2.38 ‰ was used as a standard[46]. The typical count rate of $^{12}C^-$ was $6\times10^4$-$9\times10^4$ cps. The dead time correction of the electron multipliers was applied. We carefully checked for contributions (contamination) from organic matter possibly contained in the calcite grains (ref. [47]) by monitoring $^{12}C^-$ and $^{12}C^{14}N^-$ intensities so that the presented $\delta^{13}C$ values are not compromised by contamination. The external reproducibility (2σ) of the analyses was typically about 6‰.

C-isotope analyses on the calcite and dolomite grains in Tagish Lake were performed with the NanoSIMS 50 at Max Planck Institute for Chemistry, Mainz. Because of the small grain size of the carbonate grains in Tagish Lake (typically <10 µm), we measured C isotopic ratios by recording ion images of $^{12}C^-$, $^{13}C^-$, $^{18}O^-$, $^{12}C^{14}N^-$ and $^{28}Si^-$ (3×3 μm$^2$) produced with a Cs$^+$ primary ion beam (~3 pA, about a few hundred nanometres in diameter). We calculated the $^{13}$C/$^{12}$C ratios of the carbonate grains summing up the signals from regions of interest (that is, inside of the carbonate grains, excluding the signals from the surrounding matrix). The ion images consist of 128×128 pixels and we acquired 12 sequential images for each secondary ion species. The measurement time was 9,000 μs per pixel per image, that is, about 30 min in total. Before the measurements, we used a primary ion beam with a higher current of approximately 25 pA for presputtering over 10×10 μm$^2$ areas. Examples of the ion images are shown in Supplementary Fig. 1. The same calcite standard as that for the CM carbonates was used. In addition, we newly prepared an in-house dolomite standard from Spain with a $\delta^{13}$C$_{VPDB}$ value of 3.22‰. The typical count rate of $^{12}$C$^-$ was 9×10$^3$ cps. The (minor) dead time correction of the electron multipliers was applied. Again, we carefully checked for contamination from organic matter by monitoring $^{12}$C$^-$ and $^{12}$C$^{14}$N$^-$ intensities and the homogeneity of the ion images. The presented $\delta^{13}$C values are not compromised by contamination. The external reproducibility (2σ) of the analyses was typically about 5‰.

**O isotopic ratios of calcite in Nogoya and LAP 031166.** The O isotopic ratios of the calcite grains in Nogoya and LAP 031166 measured by SIMS are shown in Fig. 1a, where the LAP 031166 data are from the literature[38]. Also shown as reference in Fig. 1a are the data of individual Ca-carbonate grains in the CM chondrites Murchison, Mighei and Cold Bokkeveld, and those obtained for whole-rock CM chondrites and their regression line from the literature[38,48]. As shown in Fig. 1a, the Nogoya and LAP 031166 data are generally consistent with the previous data: the O isotopic ratios of the CM carbonates plot along a straight line with a slope of 0.6-0.7 in an oxygen three-isotope plot, with the type 2 calcite having lower $\delta^{17,18}$O values than the type 1 calcite[36-38,48-52]. The O isotopic ratios of the carbonates likely reflect the formation temperatures and/or the isotopic ratios of water from which the carbonates precipitated. The O isotopic ratio of water is expected to have changed during aqueous alteration due to isotope exchange between anhydrous rock and water. The primordial water of CMs originally had higher $\delta^{17,18}$O and $\Delta^{17}$O values ($\Delta^{17}$O represents deviation from the terrestrial fractionation line: defined by $\Delta^{17}$O=$\delta^{17}$O–0.52×$\delta^{18}$O) (ref. [53]), while the anhydrous rock had lower values[54]. Therefore, the water composition would have changed from higher $\delta^{17,18}$O and $\Delta^{17}$O values to lower values. If correct, the type 2 calcite formed in more 'evolved' fluids than type 1. Nevertheless, both type 1 and 2 calcites show large variations in the $\delta^{13}$C values, which are not correlated with $\delta^{18}$O values. These observations suggest that the C isotopic ratios of the carbonates were not simply determined by the formation temperatures and/or the degrees of aqueous alteration, but reflect the isotopic heterogeneity of carbon sources.

**Estimate of $CO_2$/$H_2O$ ratios of ice in CM chondrites and Tagish Lake.** On the basis of the abundances and C isotopic ratios of carbonates, we inferred the abundances of $CO_2$ ice accreted to the CM and Tagish Lake parent bodies. In addition, we needed to estimate the $H_2O$ abundances of CMs and Tagish Lake to deduce the $CO_2$/$H_2O$ ratios of ice. Assuming a closed-system behaviour of $H_2O$ (the case for an open system is discussed in the Supplementary Information), one can infer the $H_2O$ abundances of chondrites from the hydrogen contents of hydrous minerals. In a previous study, the hydrogen contents of hydrous minerals in CMs were calculated by subtracting hydrogen in organic matter from the bulk hydrogen contents[19]. The hydrogen contents in organic matter were calculated from bulk carbon contents and a H/C ratio of organic matter. It is assumed that the bulk carbon is entirely present in organic matter and the H/C ratio of organic matter is represented by that of insoluble organic matter (IOM) (0.055 by weight)[55].

    Likewise, the $H_2O$ abundance of Tagish Lake was estimated from the hydrogen content of hydrous minerals; we subtracted hydrogen in organic matter from the bulk hydrogen content. The bulk hydrogen contents of three Tagish Lake fragments, namely, 11i, 11h and 5b, are 0.738, 0.872 and 0.945 wt%, respectively[18]. We deduced the hydrogen contents of organic matter using the carbon contents and H/C ratios of organic matter represented by that of IOM. We calculated the carbon contents of organic matter in the three fragments (2.65, 2.83 and 2.81 wt%, respectively) by subtracting the carbon content of the carbonate (1.3 wt%) from the bulk carbon contents (3.95, 4.13 and 4.11 wt%, respectively) reported in the literature[16,18]. The atomic H/C ratios of IOM in the three fragments were 0.51, 0.594 and 0.72, respectively[56]. Thus, the calculated hydrogen contents of hydrous minerals in the three fragments are 0.625, 0.732 and 0.776 wt%, respectively, and the average is 0.711 wt%, which corresponds to the $H_2O$ abundance of approximately 6.4 wt%.

**Uncertainty of the inferred abundances of $CO_2$ ice in CM chondrites and Tagish Lake.** In the main text, we deduced the abundances of $CO_2$ ice in CMs and Tagish Lake by mass-balance calculations, assuming that the observed C isotopic ratios of the carbonates reflect mixing between two carbon reservoirs, that is, $^{13}$C-rich $CO_2$ ice and another unknown reservoir poor in $^{13}$C. Therefore, the calculated abundances of $CO_2$ ice are dependent on the endmember compositions. As described in the main text, we assumed the $\delta^{13}$C values of the endmembers to be 80 and 20‰, and we found that ~42 and 80% of the carbonate carbon in CMs and Tagish Lake are derived from $CO_2$ ice, respectively. Instead, if we assume the $\delta^{13}C$ values of the endmembers to be 80‰ and -30‰, the latter of which is similar to the $\delta^{13}C$ values of the trapped CO and free aromatic molecules in Murchison[57,58], then we obtain about 68 and 89 % for the carbonate carbon fraction derived from $CO_2$ ice. These modifications make the inferred $CO_2/H_2O$ mole ratios of the CM ices higher by a factor of approximately 1.5, which does not change our conclusion that the CM ice has generally lower $CO_2/H_2O$ ratios than comets, whereas the $CO_2/H_2O$ ratio of the Tagish Lake ice is within the cometary range.

**Data Availability**

The data that support the plots within this paper and other findings of this study are available from the corresponding author upon reasonable request.